# Breaking the Selection Rules of Spin-Forbidden Molecular Absorption in Plasmonic Nanocavities

Oluwafemi S. Ojambati,[1] William M. Deacon,[1] Rohit Chikkaraddy,[1] Charlie Readman,[1] Qianqi Lin,[1] Zsuzsanna Koczor-Benda,[2] Edina Rosta,[2] Oren A. Scherman,[3] Jeremy J. Baumberg[1]*

[1] NanoPhotonics Centre, Cavendish Laboratory, Department of Physics, JJ Thompson Avenue, University of Cambridge, Cambridge, CB3 0HE, United Kingdom
[2] Department of Chemistry, King's College London, 7 Trinity Street, London SE1 1DB, United Kingdom
[3] Melville Laboratory for Polymer Synthesis, Department of Chemistry, University of Cambridge, Lensfield Road, Cambridge CB21EW, United Kingdom

**ABSTRACT**

Controlling absorption and emission of organic molecules is crucial for efficient light-emitting diodes, organic solar cells and single-molecule spectroscopy. Here, a new molecular absorption is activated inside a gold plasmonic nanocavity, and found to break selection rules via spin-orbit coupling. Photoluminescence excitation scans reveal absorption from a normally spin-forbidden singlet to triplet state transition, while drastically enhancing the emission rate by several thousand fold. The experimental results are supported by density functional theory, revealing the manipulation of molecular absorption by nearby metallic gold atoms.

**INTRODUCTION**

Selection rules govern absorption and emission of light in atomic and molecular systems[1], that stem from quantum mechanical symmetries dictating which atomic, electronic or vibrational transitions are allowed or forbidden[1,2]. Allowed transitions are desired for lasers and light emitting diodes, but forbidden transitions are typically inaccessible to optical excitation. Mechanisms that break these selection rules to allow forbidden transitions yield novel and efficient devices[3–6].

Molecular light emission is typically limited to an internal quantum yield of only 25% through the singlet-singlet transitions, while 75% of pathways are spin forbidden singlet-triplet transitions (Fig. 1a). To allow these forbidden transitions, spin-orbit coupling is induced by interacting the angular motion of electron spins with the magnetic dipole created by local massive atomic nuclei[1,7]. This is achieved in organo-transition-metal complexes[8–11] via *internal* heavy atom effect when a metal-to-ligand charge transfer state is formed, allowing intersystem crossing from the excited singlet state $S_1$ to excited triplet state $T_1$ (Fig. 1b). *External* heavy-atom effects induce spin mixing by placing heavy atoms near an emitter without actual bond formation[7]. Heavy atom effects have been used to enhance emission rates[12–14], thermally- and optically-activate 'delayed' fluorescence[15–18], reversibly control emission[19–21] and activate light emitting diodes (LEDs)[8,22,23]. These approaches enhance triplet emission $T_1 \rightarrow S_0$ through exciting $S_0 \rightarrow S_1$, then followed by intersystem crossing $S_1 \rightarrow T_1$ (Fig. 1b) in bulk ensembles of molecules. Due to weak spin-orbit coupling, directly accessing such forbidden singlet-triplet transition ($S_0 \rightarrow T_1$) has thus been inaccessible at the molecular level. Manipulating spin mixing to activate absorption and emission pathways at the nanoscale has however promising implications for nano-LEDs[24], nano-lasers[25], nano-solar cells[26], single molecule spectroscopy[27,28], opto-magnetism[29,30], as well as single-photon quantum emitters[31,32].

Here we activate a direct absorption from $S_0 \rightarrow T_1$ (a forbidden transition) using a nanophotonic construct that induces spin mixing to allow absorption from the forbidden transition (Fig. 1c). We employ a nanoparticle-on-mirror (NPoM) plasmonic nanocavity that achieves an extreme optical field confinement below 25 nm$^3$,[33] with fields enhanced up to 300 times in these deeply subwavelength nanogaps (Fig. 1d,e). The NPoM nanocavities each consist of an 80 nm Au nanoparticle on a Au film spaced by a monolayer of the molecular emitter Rubby [Tris(2,2'-bipyridine) ruthenium(II) hexa-fluorophosphate], with ~30 strongly emitting molecules under each Au nanoparticle[34] (for sample preparation see Supplementary Information 1). Rubby is a widely studied triplet emitter with a quantum yield of <3%[8], absorbing in the ultraviolet ~450 nm and has a large Stokes shift with a phosphorescence peak at 620 nm (Fig. 1f). The tail of this broad emission is coupled here to the NPoM cavities which have a fundamental plasmon resonance in the near- infrared at 830 ± 30 nm, set by the Rubby monolayer height which creates a gap size ~1 nm[35].

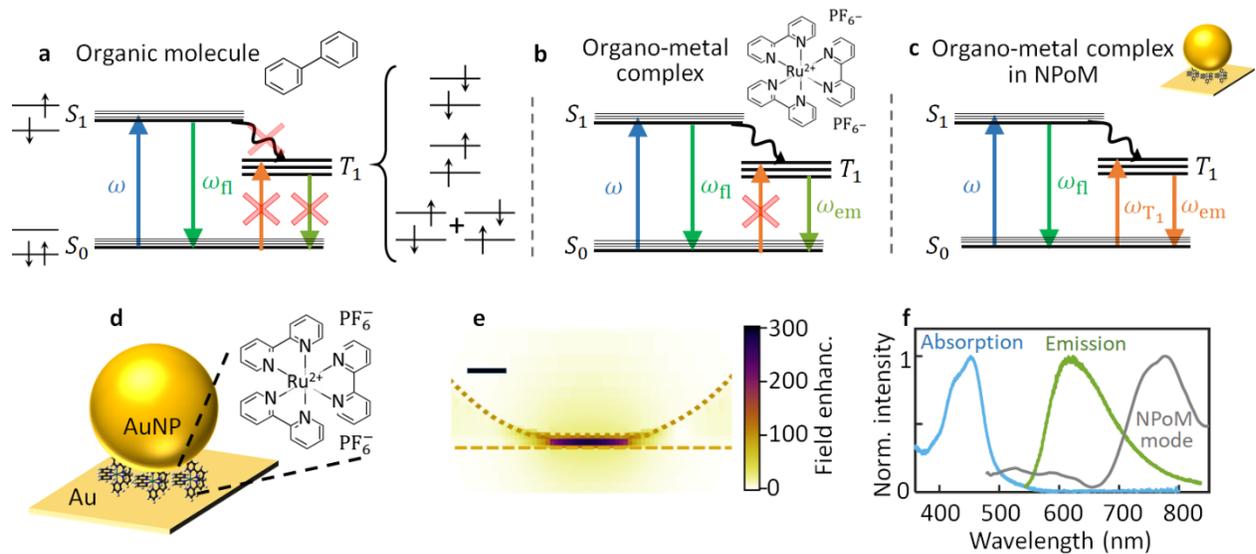

**Figure 1.** (a-c) Energy levels of allowed and forbidden transitions in molecular emitters embedded in different geometries, with (a) electron spin configurations of singlet $S_0$, $S_1$ (antiparallel electron spin pairs) and triplet $T_1$ (unpaired parallel electron spins) states. (d) Nanoparticle-on-mirror (NPoM) construct with Rubby spacer. (e) Finite-difference time domain simulation of field enhancement in the NPoM gap at $\lambda$=750 nm. Dashed lines show boundaries of Au nanoparticle and Au film, scale bar is 10 nm. (f) Absorption (blue) and emission (green) curves of Rubby in solution with 520 nm excitation. Grey curve is darkfield scattering showing dominant coupled mode of a NPoM.

**RESULTS AND DISCUSSION**

A single nanocavity is first irradiated at the forbidden transition $S_0 \rightarrow T_1$ with an excitation wavelength of 640 nm, close to the phosphorescence peak using 1 ps pulses (for experimental setup see SI 2). With average power of 1 µW on individual NPoMs, a broad spectral emission is seen with a maximum at ~700 nm and a broad tail beyond 800 nm (Fig. 2a). The broad emission has additional sharp peaks attributed to surface-enhanced resonant Raman scattering (SERRS). This emission is completely absent for Rubby in solution (80 µM, Fig.2a, solid green). To check the nature of this emission from NPoMs, the total emission intensity is found to be linearly proportional to input power (Fig. 2b). This confirms the emission comes from one-photon excitation rather than multiphoton excitation or other nonlinear processes, and no saturation is observed. Time-correlated single photon counting (TCSPC) gives the emission lifetime as $\tau_B = 520 \pm 10$ ns in bulk, which is drastically shortened to $\tau_{NPoM} < 0.2 \pm 0.1$ ns in NPoMs (Fig. 2c). While this NPoM Rubby measurement is limited by the TCSPC instrument response, it shows over three orders of magnitude reduction in spontaneous lifetime due to the high optical density of states in these nanocavities. Finite-difference time domain (FDTD) simulations reveal that there is a

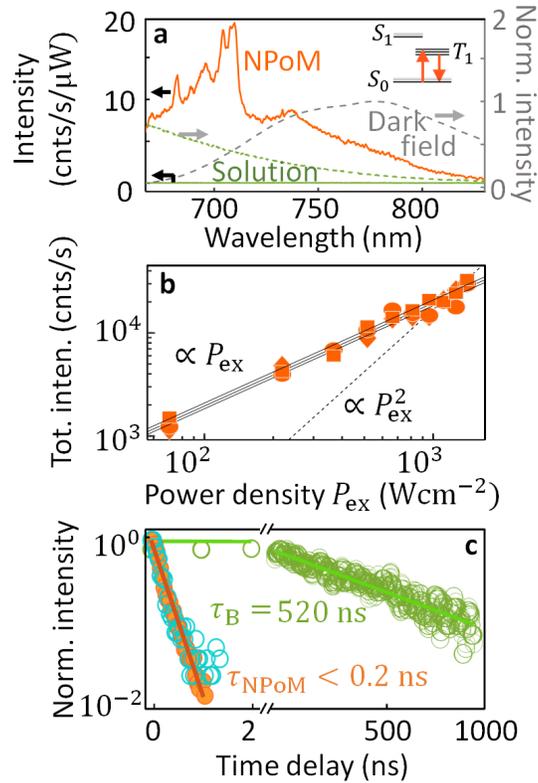

**Figure 2.** (a) Emission from Rubpy in NPoM gaps (orange) and in solution (solid green) with $\lambda_{ex}$ = 640 nm excitation. For comparison, emission from solution with 520 nm excitation (dashed green) and darkfield scattering of NPoM (dashed grey) are shown. Inset gives energy levels of excitation and emission. (b) Power dependence of emission. (c) Time-resolved emission decay in NPoM (orange), and in solution (green), with instrumental response (blue).

Purcell factor of $\sim 10^6$ for these 1 nm nanogaps, suggesting lifetimes $\sim$500 fs (for FDTD results, see SI Fig. S4). Moreover, the emission quantum yield increases more than ten-fold, to 35% for Rubpy, due to the resulting increase in radiative decay rate inside the nanocavity. Over long timescans of > 1000 seconds at 0.1 μW excitation, we observe no significant reduction in intensity which implies that these emitters are stable in NPoMs and there is no observable bleaching (SI Fig. S6).

To further probe the forbidden $S_0 \rightarrow T_1$ transition, photoluminescence excitation (PLE) pump wavelength scans are performed from $\lambda_{ex}$ = 590 - 720 nm. At each $\lambda_{ex}$, the average laser power on the sample is set to 1μW, precalibrated to account for power variations from wavelength-dependent transmission through the optical beamline. A consistent broad emission with additional SERRS peaks is seen for all $\lambda_{ex}$ (Fig. 3a), increasing and then decreasing as $\lambda_{ex}$ is increased. This is unaffected when using the scattering resonance to normalize for NPoM outcoupling efficiencies (see SI Fig. S7). The resonant absorption in PLE from the integrated emission is maximum at 642±2 nm and identical for different NPoMs (Fig.3b, average over 3 NPoMs), confirming it arises from the molecules in the gap. Note no emission is seen without Rubpy in the plasmonic gap. By contrast, the PLE of Rubpy in solution decreases steadily with $\lambda_{ex}$ (Fig. 3c), mapping the tail of the absorption line (Fig. 1e). To further confirm the general nature of our observations, we show similar results for spacers of two other organo-metallic complexes, ferrocene and a Zn porphyrin, which also give a new excitation resonance at their $S_0 \rightarrow T_1$ transitions (Fig. 3d,e, SI 8).

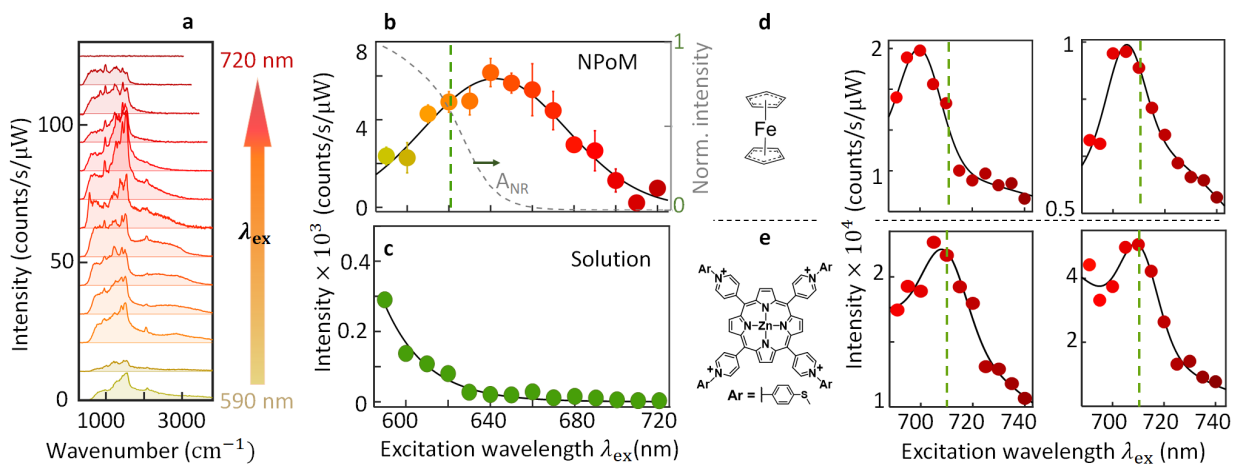

**Figure 3.** (a) Photoluminescence excitation (PLE) scan on singlet $S_0$ to triplet $T_1$ transition. (a) Emission spectra of Rubpy in NPoM *vs* energy shift from pump for increasing $\lambda_{ex}$ = 590-720 nm in 10 nm steps. (b,c) Integrated emission (PLE) spectra for NPoM *vs* Rubpy in solution, black curves are Gaussian fits. In (b), the grey dashed curve $A_{NR}$ is the predicted absorption spectrum from near-field enhancement in the NPoM gap and the dashed vertical line is the expected triplet emission energy. (d,e) Chemical structures (left panel) of ferrocene and Zn porphyrin used as NPoM spacers to obtain the integrated emission (PLE) from two NPoMs for each (middle, right panels). Dashed vertical lines mark expected triplet state energies.

In order to understand the new absorption lineshape, we perform time-dependent density functional theory (TDDFT) on $Au_2$-Rubpy-$Au_2$ to model the NPoM environment and calculate the absorption spectra for different gap sizes $d$ (Fig. 4a, for TDDFT details see SI 9). A new absorption peak appears in the presence of Au atoms that is absent in solution and which gains intensity as the gap decreases (Fig. 4b). The transitions responsible for this increase are mixed singlet-triplet transitions, induced by spin-orbit coupling. These results agree with our measurements in Fig. 3(b,c), confirming that NPoMs allow absorption at states that are spin-forbidden in solution. The sub-nm proximity of Au atoms in both facets to the emitters induces spin-orbit coupling in the molecules, thus modifying the electronic transitions and allowing direct absorption to the forbidden triplet state. At the same time, the nanocavity geometry gives efficient outcoupling without plasmonic quenching of the emission at such sub-nm distances.

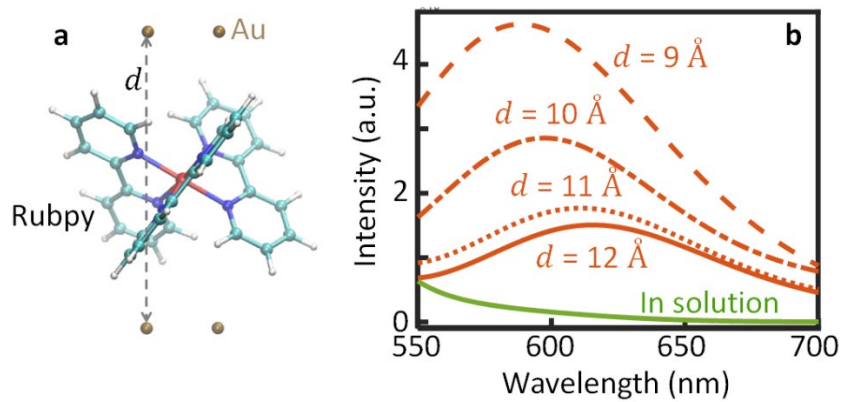

**Figure 4.** Time-dependent density functional theory simulation of the absorption of Rubpy. (a) $Au_2$-Rubpy-$Au_2$ system used for modelling the effect of the surrounding Au facets, at different gap sizes $d$. (b) Calculated absorption spectrum for solvated Rubpy and $Au_2$-Rubpy-$Au_2$ at different gap sizes $d$.

Previous theoretical studies proposed that a high field gradient in a nanocavity can induce *ac* magnetic fields that break symmetry, allowing excitation of forbidden transitions.[30,36–40] To verify if this mechanism plays a role in our observations, we calculate the spatial distribution of the magnetic field component in NPoM gaps using FDTD simulations. The magnetic field is highest at the facet edges under the nanoparticle, but zero around the centre of the gap, where molecules with the highest out-coupled emission are located (SI Fig. S5). This implies that the influence of the *ac* magnetic field on the bright molecules is minimal and thus the high field gradient effect is not responsible for the observed absorption peak. Moreover, enhanced near-field absorption ($A_{NR}$), which is calculated as the product of the absorption of Rubpy in solution and the near-field enhancement spectrum cannot explain the observed peak. The calculated $A_{NR}$ deviates significantly from the observed absorption curve (grey dashed curve, Fig. 3b). We thus identify the external heavy atom effect as the mechanism that induces the new absorption transition in the molecules.

Because of its spectral position (Fig. 3b), the observed emission at the solvated $S_0 \rightarrow T_1$ excitation is attributed to mixed single-triplet electronic transitions that produce photoluminescence (PL) and resonant Raman (SERRS). For the bare molecule, the selection rules make this transition forbidden which is why phosphorescence with a long lifetime is observed in solution (Fig. 2c), through weak spin-orbit coupling. What is unexpected is the transformation from weak phosphorescence to strong photoluminescence, while at the same time as a new strong absorption line is observed at the triplet state, observed at the few molecule level. For $S_0 \rightarrow T_1$ transitions to occur, a mechanism is required to break the electronic selection rule through spin mixing. We note that both PL and resonant Raman (or SERRS) require this same spin-mixing mechanism[41,42] to elicit the resonant lineshapes observed. Thus, the presence of SERRS in our observation is a further confirmation that selection rules have been broken.

**CONCLUSIONS**

In summary, we observe a strong singlet-triplet absorption and emission for molecules confined in these plasmonic nanocavities. The field enhancement inside the nanogaps speeds up the phosphorescence through the Purcell factor of several thousand when the mode volumes are so small compared to $\lambda^3$.[38,43] At the same time, the nanocavity induces absorption at singlet-triplet transitions by breaking the electronic selection rules via the sub-nm proximity of Au atoms. Typically, bulk metals so close to molecules quench their emission completely, but the NPoM system is different in that it enhances radiative emission. As a result the effect is seen for the first time with metallic facets. The resulting effect is to convert the phosphorescent triplet emitter into an ultrafast (< 1 ps) bright luminescent source (quantum yield ~35%). Since NPoMs allow spin-forbidden transitions to become optically accessible, this opens development of more efficient organic light emitting diodes and solar cells, optically detected magnetic resonance, as well as directly accessing triplet states for fundamental spin interactions in quantum chemistry and nanophotonics.

**ACKNOWLEDGEMENTS**
We acknowledge support from EPSRC grants EP/L027151/1, EP/R013012/1, EP/P020194/1, NanoDTC, EU H2020 829067 THOR, and ERC BioNet 757850. O.S.O acknowledges a Rubicon fellowship from the Netherlands Organisation for Scientific Research. R.C. acknowledges support from Trinity College, Cambridge.

*Corresponding author:  jjb12@cam.ac.uk

**REFERENCES**
(1) Szabo, A.; Ostlund, N. S. *Modern Quantum Chemistry: Introduction to Advanced Electronic Structure Theory*; Courier Corporation, 1996.
(2) Harris, D. C.; Bertolucci, M. D. *Symmetry and Spectroscopy: An Introduction to Vibrational and Electronic Spectroscopy*; Courier Corporation, 1978.
(3) Manchon, A.; Koo, H. C.; Nitta, J.; Frolov, S. M.; Duine, R. A. New Perspectives for Rashba Spin–Orbit Coupling. *Nat. Mater.* **2015**, *14* (9), 871–882.
(4) Li, Y.; Gecevicius, M.; Qiu, J. Long Persistent Phosphors—from Fundamentals to Applications. *Chem. Soc. Rev.* **2016**, *45* (8), 2090–2136.
(5) Soumyanarayanan, A.; Reyren, N.; Fert, A.; Panagopoulos, C. Emergent Phenomena Induced by Spin–Orbit Coupling at Surfaces and Interfaces. *Nature* **2016**, *539* (7630), 509–517.
(6) Reineke, S.; Thomschke, M.; Lüssem, B.; Leo, K. White Organic Light-Emitting Diodes: Status and Perspective. *Rev. Mod. Phys.* **2013**, *85* (3), 1245–1293.
(7) McGlynn, S. P.; Azumi, T.; Kinoshita, M. *Molecular Spectroscopy of the Triplet State*; Prentice-Hall, 1969.
(8) Yersin, H.; Rausch, A. F.; Czerwieniec, R.; Hofbeck, T.; Fischer, T. The Triplet State of Organo-Transition Metal Compounds. Triplet Harvesting and Singlet Harvesting for Efficient OLEDs. *Coord. Chem. Rev.* **2011**, *255* (21–22), 2622–2652.
(9) Baryshnikov, G.; Minaev, B.; Ågren, H. Theory and Calculation of the Phosphorescence Phenomenon. *Chem. Rev.* **2017**, *117* (9), 6500–6537.
(10) Amemori, S.; Sasaki, Y.; Yanai, N.; Kimizuka, N. Near-Infrared-to-Visible Photon Upconversion Sensitized by a Metal Complex with Spin-Forbidden yet Strong S0–T1 Absorption. *J. Am. Chem. Soc.* **2016**, *138* (28), 8702–8705.
(11) Lamansky, S.; Djurovich, P.; Murphy, D.; Abdel-Razzaq, F.; Lee, H.-E.; Adachi, C.; Burrows, P. E.; Forrest, S. R.; Thompson, M. E. Highly Phosphorescent Bis-Cyclometalated Iridium Complexes: Synthesis, Photophysical Characterization, and Use in Organic Light Emitting Diodes. *J. Am. Chem. Soc.* **2001**, *123* (18), 4304–4312.
(12) Einzinger, M.; Zhu, T.; Silva, P. de; Belger, C.; Swager, T. M.; Voorhis, T. V.; Baldo, M. A. Shorter Exciton Lifetimes via an External Heavy-Atom Effect: Alleviating the Effects of Bimolecular Processes in Organic Light-Emitting Diodes. *Adv. Mater.* **2017**, *29* (40), 1701987.
(13) Vosch, T.; Antoku, Y.; Hsiang, J.-C.; Richards, C. I.; Gonzalez, J. I.; Dickson, R. M. Strongly Emissive Individual DNA-Encapsulated Ag Nanoclusters as Single-Molecule Fluorophores. *Proc. Natl. Acad. Sci.* **2007**, *104* (31), 12616–12621.

(14) Bolton, O.; Lee, K.; Kim, H.-J.; Lin, K. Y.; Kim, J. Activating Efficient Phosphorescence from Purely Organic Materials by Crystal Design. *Nat. Chem.* **2011**, *3* (3), 205–210.
(15) Fleischer, B. C.; Petty, J. T.; Hsiang, J.-C.; Dickson, R. M. Optically Activated Delayed Fluorescence. *J. Phys. Chem. Lett.* **2017**, *8* (15), 3536–3543.
(16) Evans, E. W.; Olivier, Y.; Puttisong, Y.; Myers, W. K.; Hele, T. J. H.; Menke, S. M.; Thomas, T. H.; Credgington, D.; Beljonne, D.; Friend, R. H.; Greenham, N. C. Vibrationally Assisted Intersystem Crossing in Benchmark Thermally Activated Delayed Fluorescence Molecules. *J. Phys. Chem. Lett.* **2018**, *9* (14), 4053–4058.
(17) Yang, Z.; Mao, Z.; Xie, Z.; Zhang, Y.; Liu, S.; Zhao, J.; Xu, J.; Chi, Z.; Aldred, M. P. Recent Advances in Organic Thermally Activated Delayed Fluorescence Materials. *Chem. Soc. Rev.* **2017**, *46* (3), 915–1016.
(18) Leitl, M. J.; Krylova, V. A.; Djurovich, P. I.; Thompson, M. E.; Yersin, H. Phosphorescence versus Thermally Activated Delayed Fluorescence. Controlling Singlet–Triplet Splitting in Brightly Emitting and Sublimable Cu(I) Compounds. *J. Am. Chem. Soc.* **2014**, *136* (45), 16032–16038.
(19) Mieno, H.; Kabe, R.; Adachi, C. Reversible Control of Triplet Dynamics in Metal-Organic Framework-Entrapped Organic Emitters via External Gases. *Commun. Chem.* **2018**, *1* (1), 27.
(20) Downing, C. A.; Carreño, J. C. L.; Laussy, F. P.; del Valle, E.; Fernández-Domínguez, A. I. Quasichiral Interactions between Quantum Emitters at the Nanoscale. *Phys. Rev. Lett.* **2019**, *122* (5), 057401.
(21) Eizner, E.; Martínez-Martínez, L. A.; Yuen-Zhou, J.; Kéna-Cohen, S. Inverting Singlet and Triplet Excited States Using Strong Light-Matter Coupling. *Sci. Adv.* **2019**, *5* (12), eaax4482.
(22) Shizu, K.; Uejima, M.; Nomura, H.; Sato, T.; Tanaka, K.; Kaji, H.; Adachi, C. Enhanced Electroluminescence from a Thermally Activated Delayed-Fluorescence Emitter by Suppressing Nonradiative Decay. *Phys. Rev. Appl.* **2015**, *3* (1), 014001.
(23) Lai, P.-N.; Brysacz, C. H.; Alam, M. K.; Ayoub, N. A.; Gray, T. G.; Bao, J.; Teets, T. S. Highly Efficient Red-Emitting Bis-Cyclometalated Iridium Complexes. *J. Am. Chem. Soc.* **2018**, *140* (32), 10198–10207.
(24) Lozano, G.; Rodriguez, S. R.; Verschuuren, M. A.; Gómez Rivas, J. Metallic Nanostructures for Efficient LED Lighting. *Light Sci. Appl.* **2016**, *5* (6), e16080–e16080.
(25) Zhou, W.; Dridi, M.; Suh, J. Y.; Kim, C. H.; Co, D. T.; Wasielewski, M. R.; Schatz, G. C.; Odom, T. W. Lasing Action in Strongly Coupled Plasmonic Nanocavity Arrays. *Nat. Nanotechnol.* **2013**, *8* (7), 506–511.
(26) Atwater, H. A.; Polman, A. Plasmonics for Improved Photovoltaic Devices. *Nat. Mater.* **2010**, *9* (3), 205–213.
(27) Kukura, P.; Celebrano, M.; Renn, A.; Sandoghdar, V. Single-Molecule Sensitivity in Optical Absorption at Room Temperature. *J. Phys. Chem. Lett.* **2010**, *1* (23), 3323–3327.
(28) Moerner, W. E.; Shechtman, Y.; Wang, Q. Single-Molecule Spectroscopy and Imaging over the Decades. *Faraday Discuss.* **2015**, *184* (0), 9–36.
(29) Balasubramanian, G.; Chan, I. Y.; Kolesov, R.; Al-Hmoud, M.; Tisler, J.; Shin, C.; Kim, C.; Wojcik, A.; Hemmer, P. R.; Krueger, A.; Hanke, T.; Leitenstorfer, A.; Bratschitsch, R.; Jelezko, F.; Wrachtrup, J. Nanoscale Imaging Magnetometry with Diamond Spins under Ambient Conditions. *Nature* **2008**, *455* (7213), 648–651.
(30) Duan, S.; Rinkevicius, Z.; Tian, G.; Luo, Y. Optomagnetic Effect Induced by Magnetized Nanocavity Plasmon. *J. Am. Chem. Soc.* **2019**, *141* (35), 13795–13798..
(31) Siampour, H.; Kumar, S.; Bozhevolnyi, S. I. Nanofabrication of Plasmonic Circuits Containing Single Photon Sources. *ACS Photonics* **2017**, *4* (8), 1879–1884.
(32) Singh, A.; de Roque, P. M.; Calbris, G.; Hugall, J. T.; van Hulst, N. F. Nanoscale Mapping and Control of Antenna-Coupling Strength for Bright Single Photon Sources. *Nano Lett.* **2018**, *18* (4), 2538–2544.

(33) Baumberg, J. J.; Aizpurua, J.; Mikkelsen, M. H.; Smith, D. R. Extreme Nanophotonics from Ultrathin Metallic Gaps. *Nat. Mater.* **2019**, *18*, 668.

(34) O. S. Ojambati; et al. Enhanced Two-Photon Absorption in Plasmonic Nanocavities. *arXiv.org* **2019**.

(35) de la Llave, E.; Herrera, S. E.; Méndez De Leo, L. P.; Williams, F. J. Molecular and Electronic Structure of Self-Assembled Monolayers Containing Ruthenium(II) Complexes on Gold Surfaces. *J. Phys. Chem. C* **2014**, *118* (37), 21420–21427.

(36) Zurita-Sánchez, J. R.; Novotny, L. Multipolar Interband Absorption in a Semiconductor Quantum Dot. I. Electric Quadrupole Enhancement. *JOSA B* **2002**, *19* (6), 1355–1362.

(37) Filter, R.; Mühlig, S.; Eichelkraut, T.; Rockstuhl, C.; Lederer, F. Controlling the Dynamics of Quantum Mechanical Systems Sustaining Dipole-Forbidden Transitions via Optical Nanoantennas. *Phys. Rev. B* **2012**, *86* (3), 035404.

(38) Rivera, N.; Kaminer, I.; Zhen, B.; Joannopoulos, J. D.; Soljačić, M. Shrinking Light to Allow Forbidden Transitions on the Atomic Scale. *Science* **2016**, *353* (6296), 263–269.

(39) Bossini, D.; Belotelov, V. I.; Zvezdin, A. K.; Kalish, A. N.; Kimel, A. V. Magnetoplasmonics and Femtosecond Optomagnetism at the Nanoscale. *ACS Photonics* **2016**, *3* (8), 1385–1400.

(40) Neuman, T.; Esteban, R.; Casanova, D.; García-Vidal, F. J.; Aizpurua, J. Coupling of Molecular Emitters and Plasmonic Cavities beyond the Point-Dipole Approximation. *Nano Lett.* **2018**, *18* (4), 2358–2364.

(41) Köppen, T.; Franz, D.; Schramm, A.; Heyn, Ch.; Heitmann, D.; Kipp, T. Resonant Raman Transitions into Singlet and Triplet States in InGaAs Quantum Dots Containing Two Electrons. *Phys. Rev. Lett.* **2009**, *103* (3), 037402.

(42) Sweeney, T. M.; Carter, S. G.; Bracker, A. S.; Kim, M.; Kim, C. S.; Yang, L.; Vora, P. M.; Brereton, P. G.; Cleveland, E. R.; Gammon, D. Cavity-Stimulated Raman Emission from a Single Quantum Dot Spin. *Nat. Photonics* **2014**, *8* (6), 442–447.

(43) Akselrod, G. M.; Argyropoulos, C.; Hoang, T. B.; Ciracì, C.; Fang, C.; Huang, J.; Smith, D. R.; Mikkelsen, M. H. Probing the Mechanisms of Large Purcell Enhancement in Plasmonic Nanoantennas. *Nat. Photonics* **2014**, *8* (11), 835–840.